\documentclass[twocolumn,aps,longbibliography]{revtex4-2}

\usepackage{bm}
\usepackage{graphicx}
\usepackage{colordvi}
\usepackage{color}

\definecolor{xred}{rgb}{1,0,0}
\definecolor{xblue}{rgb}{0,0,1}

\graphicspath{{../},{figures/}}

\begin{document}


\title{Quantumness and its hierarchies in PT-symmetric down-conversion models}

\author{Jan Pe\v{r}ina Jr.}
\email{jan.perina.jr@upol.cz} \affiliation{Joint Laboratory of
Optics, Faculty of Science, Palack\'{y} University, 17. listopadu
12, 771~46 Olomouc, Czech Republic}
\affiliation{Institute of
Physics of the Academy of Sciences of the Czech Republic, Joint
Laboratory of Optics of Palack\'{y} University and Institute of
Physics AS CR, 17. listopadu 50a, Olomouc 772 07, Czech Republic}

\author{Karol Bartkiewicz}
\affiliation{Institute of Spintronics and Quantum Information,
Faculty of Physics and Astronomy, Adam Mickiewicz University,
61-614 Pozna\' n, Poland}

\author{Grzegorz Chimczak}
\affiliation{Institute of Spintronics and Quantum Information,
Faculty of Physics and Astronomy, Adam Mickiewicz University,
61-614 Pozna\' n, Poland}

\author{Anna Kowalewska-Kudlaszyk}
\affiliation{Institute of Spintronics and Quantum Information,
Faculty of Physics and Astronomy, Adam Mickiewicz University,
61-614 Pozna\' n, Poland}

\author{Adam Miranowicz}
\affiliation{Institute of Spintronics and Quantum Information,
Faculty of Physics and Astronomy, Adam Mickiewicz University,
61-614 Pozna\' n, Poland}

\author{Joanna K. Kalaga}
\affiliation{Quantum Optics and Engineering Division, Faculty of
Physics and Astronomy, University of Zielona G\' ora, Prof. Z.
Szafrana 4a, 65-516 Zielona G\' ora, Poland}

\author{Wies\l aw Leo\' nski}
\affiliation{Quantum Optics and Engineering Division, Faculty of
Physics and Astronomy, University of Zielona G\' ora, Prof. Z.
Szafrana 4a, 65-516 Zielona G\' ora, Poland}

\date{\today}

\begin{abstract}
We investigate the hierarchy of quantum correlations in a
quadratic bosonic parity-time-symmetric system (PTSS) featuring
distinct dissipation and amplification channels. The hierarchy
includes global nonclassicality, entanglement, asymmetric quantum
steering, and Bell nonlocality. We elucidate the interplay between
the system  {physical} nonlinearity---which serves as a source of
quantumness---and the specific dynamics of bosonic PTSSs, which
are qualitatively influenced by their damping and amplification
characteristics. Using a set of quantifiers---including local and
global nonclassicality depths, negativity, steering parameters,
and the Bell parameter---we demonstrate that the standard PTSS
typically exhibits weaker quantumness than its counterparts
affected solely by damping or solely by amplification. Both the
maximum values attained by these quantifiers and the speed and
duration of their generation are generally lower in the standard
PTSS. A comparative analysis of three two-mode PTSSs---standard,
passive, and active---with identical eigenvectors and real parts
of eigenfrequencies, but differing in their damping and
amplification strengths, reveals the crucial role of quantum
fluctuations associated with gain and loss. Among them, the
passive PTSS yields the most strongly nonclassical states.
Nevertheless, under suitable conditions, the standard PTSS can
also generate highly nonclassical states. The supremacy of the
passive PTSS is further supported by its fundamental advantages in
practical realizations.
\end{abstract}

\maketitle

\section{Introduction}

Since the seminal works on parity-time-symmetric systems (PTSSs)
by Bender \emph{et al.}~\cite{Bender1998,Bender1999},
non-Hermitian Hamiltonians with real eigenvalues have attracted
considerable attention in the physics community~\cite{Bender2003}.
This interest stems from the fact that PTSSs possess unique
structures in their Hilbert (or Liouville) spaces, which exhibit
degeneracies---both in eigenvalues and eigenvectors---at specific
parameter values known as exceptional points (EPs). These
degeneracies give rise to a variety of intriguing physical
phenomena (for details, see reviews~\cite{Ozdemir2019,Miri2019}).
Systems operating at or near EPs can be harnessed for a range of
applications, including enhanced
sensing~\cite{Liu2016,Chen2017,Hodaei2017}, enhanced nonlinear
interactions~\cite{He2015,Vashahri2017,PerinaJr2019x,PerinaJr2019y},
unidirectional light propagation~\cite{Peng2014,Chang2014}, and
even invisibility cloaking~\cite{Lin2011,Regen2012}.

Interesting behaviors of nonlinear optical systems have also been
studied under PTSS conditions, i.e., when damping and
amplification are balanced. It has been shown that highly
nonclassical states can be generated in such systems under
suitable conditions~\cite{He2015, Vashahri2017, PerinaJr2019x,
PerinaJr2019y}, despite the unavoidable noise present in quantum
systems involving damping and/or amplification~\cite{Scheel2018}.
This raises an important question: to what extent does the
specific dynamics in the Hilbert space of bosonic PTSSs influence
the ability of system  {physical} nonlinearities  {(though
analyzed in many cases in their linearized versions)} to generate
quantumness~\cite{Boyd2003,Mandel1995,Perina1991,PerinaJr2000}?
The system's dynamics affect the rate at which different forms of
quantumness emerge, the maximal values attained by various
quantifiers, and their asymptotic limits. A fundamental problem
thus arises: \emph{Can the specific dynamics in the Hilbert space
of a PTSS enhance the ability of  {physical} nonlinearities to
generate diverse forms of nonclassical states?}

 {In this paper, we address in detail this complex issue by the
extended numerical analysis of different versions of two-mode
bosonic $\mathcal{PT}$-symmetric system with parametric
down-conversion that covers the whole system's parameter space and
includes all common forms of quantumness (system nonclassicality,
entanglement, steering, the Bell nonlocality). Emergence of
quantumness is discussed as competition between the system's
coherent dynamics and detrimental influence of the reservoir
fluctuating forces. The obtained complete numerical analysis
allows us to draw even several general conclusions. Whereas the
coherent dynamics does not allow to fully compensate for the
influence of the reservoir forces for the most of the parameters
in the standard PTSS, it leads to the generation of the states
with high levels of quantumness in the passive PTSS.}

The paper is structured as follows.  {In Sec.~II we discuss the
general behavior of bosonic $\mathcal{PT}$-symmetric systems whose
properties emerge in the competition between their coherent
evolution and the influence of the noise inevitably accompanying
damping and amplification present in the system.} In Sec.~III, the
considered PTSS is introduced, its solution is found, and
statistical properties of its modes are described considering
Gaussian fields. Nonclassicality depths, negativity, steering
parameter and the Bell parameter are introduced and determined for
the Gaussian fields in Sec.~IV. The role of standard
$\mathcal{PT}$-symmetry in nonclassical-state generation is
elucidated in Sec.~V using the comparison with the systems
influenced only either by damping or amplification. In Sec.~VI,
relying on similarity of coherent dynamics in the standard PTSS,
passive PTSS with doubled damping, and active PTSS system with
doubled amplification [see the scheme in Figs.~\ref{fig1}(a,d,e)],
the role of reservoir fluctuations accompanying damping and
amplification in nonclassical-state generation is elucidated.
Section~VII is devoted to the comparison of the PTSSs ability to
generate different forms of quantumness and quantum correlations.
Time and speed aspects of the nonclassical-state generation are
addressed in Sec.~VIII. Conclusions are drawn in Sec.~IX.

\section{ {Competing effects of coherent dynamics and reservoir noise in quantum
$\mathcal{PT}$-symmetric systems}}

When we look back at the history, PTSSs were extensively studied
within the framework of classical physics (see books
\cite{Christodoulides2018book, Bender2020book} and reviews
\cite{Konotop2016, Feng2018, Ozdemir2019, Gupta2020,
ElGanainy2021})---particularly in optics, where their specific
coherent dynamics proved especially beneficial. A hallmark of
PTSSs is the simplification of system dynamics at EPs, often
accompanied by the enhancement of certain system properties. The
extension of these concepts to quantum optical bosonic systems,
via the use of Glauber coherent states and the Glauber-Sudarshan
$P$-representation of the statistical operator~\cite{Glauber1963,
Sudarshan1963}, appeared straightforward. The idea of employing
PTSSs endowed with some form of  {physical} nonlinearity to
generate various nonclassical and entangled states promised
significant and attractive outcomes. Indeed, a range of nonlinear
PTSSs have been used to produce nonclassical states with unusual
properties~\cite{He2015, Vashahri2017, PerinaJr2019x,
PerinaJr2019y}.

However, a critical problem was identified: the presence of
chaotic fluctuating forces, which---according to both the
fluctuation-dissipation and analogous amplification-fluctuation
theorems---inevitably accompany damping and amplification. Due to
their chaotic nature, these forces tend to degrade all forms of
system quantumness~\cite{Scheel2018}. As a result, two competing
effects come into play in PTSSs with respect to the generation of
quantumness: while the coherent $\mathcal{PT}$-symmetric dynamics
tends to support and enhance quantumness, the accompanying chaotic
fluctuations tend to suppress it~\cite{PerinaJr2023}. This raises
a fundamental question: \emph{Can the coherent
$\mathcal{PT}$-symmetric dynamics fully compensate for the
detrimental effects of the fluctuating forces---or even enhance
the system's quantumness despite their presence?}

While it is difficult---if not impossible---to answer this
question in full generality, valuable physical insight can be
gained by analyzing specific, well-defined models. One of the
simplest models that satisfies the necessary criteria involves two
interacting bosonic modes, mutually coupled through both linear
and  {(physically)} nonlinear interactions and subject to damping
and amplification.  {An important technical advantage of this
model lies in the fact that, under the physically relevant
conditions, the model can be analyzed in its linearized version in
which its dynamical operator equations remain linear, enabling a
fully analytical treatment.}

By focusing on Gaussian states, we are able to analytically
determine all relevant parameters characterizing the system,
including the effects of averaging over the chaotic fluctuating
forces. This analytical approach makes it possible to
systematically explore the entire parameter space of the model, as
well as to examine its temporal evolution. The study of this
evolution is supported by both numerical simulations and
asymptotic (long-time) analytical formulas.

Regarding the effect of noise in our system, it originates from
quantum sources, as it arises from interactions with quantum
reservoirs composed of two-level atoms. These atoms are either in
the ground state (in the case of damping) or in the excited state
(for amplification). The quantum state of the reservoir atoms
fundamentally alters the character of the reservoir-induced noise:
excited atoms can induce both spontaneous and stimulated emission
in the system, whereas ground-state atoms only allow for
stimulated absorption. This distinction leads to the fact that
amplification-related noise is more detrimental than
damping-related noise---it is simply stronger~\cite{Scheel2018}.

However, as already mentioned above, the generation of quantum
features in the system is governed by more than just the noise
level; it also strongly depends on the system's  {physical}
nonlinearity, which is determined by the product of the nonlinear
coupling constant and the mode amplitudes. The coherent part of
the dynamics typically leads to larger mode amplitudes, especially
in the presence of amplification, which can in turn enhance the
generation of quantumness. This stands in contrast to the
detrimental effects introduced by noise. As a result, the system's
behavior emerges from the competition between these two opposing
factors: chaotic noise versus coherent  {physically-nonlinear}
dynamics.

This interplay lies at the heart of our investigation and
constitutes the central motivation behind it. While noise often
dominates and at least partially suppresses quantum effects, there
exist regions in the parameter space where coherent dynamics
prevail, enabling the emergence of nonclassical behavior despite
the presence of noise. This delicate balance is what makes the
results fundamentally interesting and potentially attractive.

The selection of an appropriate reference system is an important
issue. The hallmark features of PTSSs arise from a balance between
gain and loss in the modes, and these features should be absent in
any suitable reference system. Such reference configurations
include systems in which only one mode is subject to damping while
the other evolves freely, or vice versa---one mode is amplified
while the other remains unaffected. However, detailed analysis
across the full parameter space reveals that, in most cases, the
$\mathcal{PT}$-symmetric dynamics does not offer a significant
advantage in generating different forms of quantumness. This
suggests that the influence of chaotic fluctuating forces is
typically too strong to be compensated for by coherent
$\mathcal{PT}$-symmetric evolution.

This naturally leads us to consider more general
$\mathcal{PT}$-symmetric-like systems---namely, their passive and
active variants~\cite{Chimczak2023}. In passive (active)
$\mathcal{PT}$-symmetric systems, the modes experience unequal
levels of damping (amplification). However, the eigenvalues of
their corresponding dynamical matrices share a common damping
(amplification) component, while the remaining parts---along with
the associated eigenvectors---retain the essential characteristics
of standard PTSSs. As such, aside from the global damping (or
amplification) factor, the coherent dynamics remain effectively
identical to those of the standard PTSS counterpart. This
structural similarity is promising for the generation of
quantumness, provided that the impact of chaotic fluctuations is
sufficiently reduced.

Indeed, the contributions of chaotic fluctuating forces are not
symmetric: those associated with damping are generally weaker than
those arising from amplification. This is because amplification
typically involves coupling to inverted two-level atoms, which are
susceptible to spontaneous photon emission---a process that
significantly enhances the destructive influence of noise. Based
on this observation, we analyze a passive PTSS in which one mode
is subject to double damping, while the other evolves freely. As
demonstrated below, this configuration turns out to be optimal for
the generation of quantumness.

Nevertheless, active PTSSs---where one mode is doubly amplified
and the other evolves without gain or loss---should not be
dismissed a priori. Despite the stronger fluctuating forces
inherent to amplification, the shared amplification factor can
significantly enhance the system's  {physical} nonlinearity
through its influence on the coherent
$\mathcal{PT}$-symmetric-like dynamics. While under typical
conditions the detrimental effects of noise dominate, our results
reveal specific parameter regimes in which amplified coherent
dynamics prevail, leading to enhanced nonclassicality, as
discussed in detail below.

In the paper, we analyze a two-mode bosonic system governed by a
quadratic Hamiltonian~\cite{PerinaJr2019x, PerinaJr2019y,
PerinaJr2023}---   {the original nonlinear Hamiltonian belonging
to the three-mode optical nonlinear interaction is linearized by
assuming a strong un-depleted pump mode (parametric approximation)
\cite{Boyd2003,Mandel1995}. This linearized Hamiltonian includes
both linear mode coupling and nonlinear interaction arising from
parametric down-conversion. Though the nonlinear interaction is
effectively described by its linearized form it enables the
generation of quantum states.} The quadratic form of the
Hamiltonian yields linear Heisenberg equations of motion, which
can be solved analytically \cite{PerinaJr2000}. These solutions
incorporate fluctuating Langevin noise operators associated with
damping and amplification, providing a rigorous framework for the
system's dynamical analysis.

 {We note that we refer to the investigated model as
(physically) nonlinear because the Hamiltonian in Eq.~(\ref{1})
below contains the terms $a_1 a_2 + a_1^\dagger a_2^\dagger$,
which correspond to the two-mode squeezing interaction that
belongs to the group of three-mode optical nonlinear interactions
described by second-order susceptibilities $\chi^{(2)}$. These
interactions are capable of generating or enhancing
nonclassicality and increasing the total number of excitations
during evolution, in contrast to linear optical systems
characterized solely by first-order susceptibilities $\chi^{(1)}$.
This fundamental distinction then underpins the definition of
certain nonclassicality measures, such as potentials of quantum
entanglement, steering, and Bell nonlocality~\cite{Kadlec2024}.}

The approach based on linearizing the nonlinear-system dynamics
and subsequent analytical treatment allow for a detailed
comparison between the standard PTSS---affected by both damping
and amplification---and two related configurations involving only
damping or only amplification [see Figs.~\ref{fig1}(a-c)]. By
evaluating various quantifiers of quantumness---characterizing
both nonclassicality and quantum correlations---we investigate how
the system's specific structure impacts the formation of
nonclassical states. The analyzed quantifiers include local and
global nonclassicality depths~\cite{Lee1991},
negativity~\cite{Hill1997, Horodecki2009} as a measure of
entanglement, the steering parameter~\cite{Cavalcanti2009} for
asymmetric quantum steering, and the Bell (nonlocality)
parameter~\cite{Bell1964} for identifying the strongest type of
quantum correlations among those considered.

We note that, in our analysis, we consider the so-called
\emph{quantum exceptional points} that occur in the dynamics of
open quantum systems characterized by their Liouvillians. Contrary
to the usual Hamiltonian EPs, they are fully compatible with the
general dynamics of open quantum systems
\cite{Minganti2019,PerinaJr2022,Thapliyal2024,PerinaJr2024}. We
note that, rather than analyzing the system's evolution via the
master equation and its associated Liouvillian, we employ a more
tractable approach based on the analytical solution of the
corresponding Langevin-Heisenberg equations, including their
Langevin noise terms \cite{Vogel2001,Perina1991}. This method not
only reveals the system's eigenfrequencies (associated with the
Liouvillian spectrum, cf. \cite{PerinaJr2022}), but also
facilitates the derivation of Gaussian-state parameters through
averaging over the chaotic noise forces.

\section{Two-mode bosonic system with damping and amplification}

\begin{figure*}[t]  
 \centerline{ \includegraphics[width=0.19\hsize]{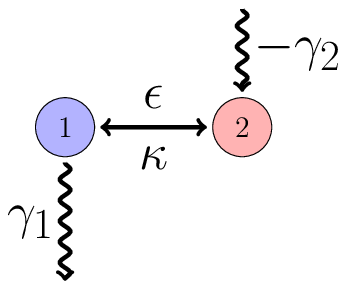}
  \hspace{4mm}
 \includegraphics[width=0.13\hsize]{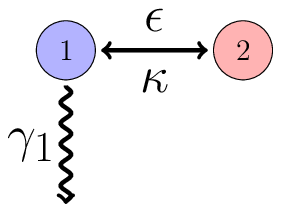}
  \hspace{4mm}
 \includegraphics[width=0.15\hsize]{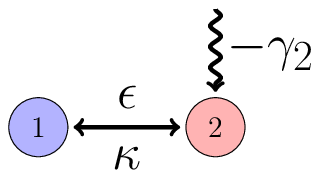}
  \hspace{4mm}
 \includegraphics[width=0.17\hsize]{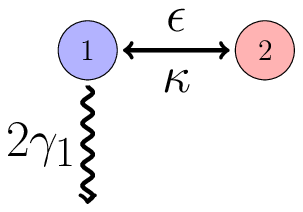}
  \hspace{4mm}\includegraphics[width=0.19\hsize]{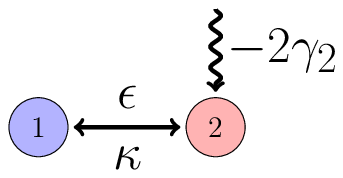}}
  \centerline{ \small (a) \hspace{.16\hsize} (b) \hspace{.16\hsize} (c)
   \hspace{.16\hsize} (d)  \hspace{.16\hsize} (e)}
 \vspace{4mm}
 \caption{Schematics of the 2-mode bosonic system analyzed under different conditions: (a) {\it standard $\mathcal{PT}$-symmetric system},
  where mode 1 is damped ($\gamma_1 > 0$) and mode 2 is amplified ($\gamma_2 < 0$);
 (b) system with only mode 1 damped ($\gamma_1 > 0$, $\gamma_2 = 0$);
 (c) system with only mode 2 amplified ($\gamma_2 < 0$, $\gamma_1 = 0$);
 (d) {\it passive $\mathcal{PT}$-symmetric system} with mode 1 doubly damped ($2\gamma_1 > 0$, $\gamma_2 = 0$);
 (e) {\it active $\mathcal{PT}$-symmetric system} with mode 2 doubly amplified ($2\gamma_2 < 0$, $\gamma_1 = 0$).
  Various parameters calculated for these systems are consistently distinguished in the following figures using the superscripts: (a)
  ad, (b) d, (c) a, (d) dd, and (e) aa.}
 \label{fig1}
\end{figure*}

The modes are assumed to mutually interact via the linear exchange
of energy (described by linear coupling constant  $ \epsilon $)
and the  {physically} nonlinear addition or subtraction of photon
pairs into both modes that originates in parametric
down-conversion \cite{Mandel1995} (nonlinear coupling constant $
\kappa $). Damping and amplification of modes, that causes the
presence of additional Langevin fluctuating operator forces
representing the back-action of the reservoir, describe loss and
addition of energy into the modes (for the sketch of the system
analyzed under different conditions, see Fig.~\ref{fig1}).
Introducing the photon annihilation ($ \hat{a}_j $) and creation
($ \hat{a}_j^\dagger $) operators of the modes denoted as 1 and 2
together with the corresponding Langevin operator forces $
\hat{l}_j $ and $ \hat{l}_j^\dagger $ and system-reservoir
coupling constants $ r_j $, $ j=1,2 $, we express the appropriate
system interaction Hamiltonian $ \hat{H} $ as
follows~\cite{Perina1991,PerinaJr2022,PerinaJr2023}:
\begin{eqnarray}  
 \hat{H} = \left[ \epsilon \hat{a}_1^\dagger\hat{a}_2 +     
   \kappa \hat{a}_1\hat{a}_2 + {\rm h.c.} \right] 
   + \left[  r_1 \hat{a}_1 \hat{l}_1^\dagger + r_2 \hat{a}_2 \hat{l}_2^\dagger
   + {\rm h.c.} \right] ,
\label{1}
\end{eqnarray}
where symbol h.c. replaces the Hermitian conjugated terms. To
allow consistent description of both damped and amplified modes,
we chose the Langevin operator forces $ \hat{l}_j $ and $
\hat{l}_j^\dagger $, $ j=1,2 $, as the Pauli spin-flip operators
that describe the reservoir two-level atoms \cite{Meystre2007}.
Invoking the second-order perturbation theory in the
system-reservoir coupling constants $ r_1 $ and $ r_2 $ and
eliminating the reservoir operators by replacing them by their
reservoir mean values [with the two-level reservoir atoms in the
ground (excited) state for damping (amplification)] we reveal the
corresponding damping and amplification constants as well as the
appropriate mean values of the Langevin operator forces, see
Eqs.~(\ref{4}) and (\ref{2}) below as well as detailed derivation
in Ref.~\cite{PerinaJr2022,Vogel2001,Perina1991}.

We note that the above approach based on the Heisenberg-Langevin
operator equations with the Langevin fluctuating operator forces
\cite{Perina1991} represents an alternative to the commonly used
approach based on the (generalized) master equation for a
statistical operator \cite{Vogel2001,Jager2022}. Convenience of
the application of these approaches differ according to the
situation that includes both the structure of the model and the
type of the states investigated. In our case in which we pay
attention to Gaussian fields \cite{Adesso2007,Perina1991}, the use
of the Heisenberg-Langevin operator equations is more convenient
owing to their linearity  {that arises in the parametric
approximation applied to the nonlinear interaction}
\cite{PerinaJr2000}. In contrast, the solution of the
corresponding master equation transformed into the form of the
Fokker-Planck equation \cite{Risken1989} would involve the
temporal solution for the mean-field-operator amplitudes and also
statistical coefficients of the Gaussian states [see
Eq.~(\ref{20}) below]. In the applied approach, these coefficients
are derived directly from the operator solution of the
Heisenberg-Langevin equations. This considerably simplifies the
calculations.

In the model, we assume that mode 1 is damped (damping constant $
\gamma_1 $ whereas mode 2 is amplified (amplification constant $
-\gamma_2 $. The reservoirs responsible for damping and
amplification are assumed to be described by independent quantum
random Gaussian and Markovian processes with the following
characteristics~\cite{Meystre2007,Agarwal2012,Perinova2019}, $
j=1,2 $:
\begin{eqnarray} 
 & \langle\hat{l}_j(t)\rangle = \langle\hat{l}_j^\dagger(t)\rangle = 0,& \nonumber \\
 &\langle\hat{l}_j^\dagger(t)\hat{l}_j(t')\rangle = \tilde{l}_j\delta(t-t'),& \nonumber \\
 & \langle\hat{l}_j(t)\hat{l}_j^\dagger(t')\rangle =
  l_j \delta(t-t').&
\label{2}
\end{eqnarray}
In Eq.~(\ref{2}), the real constants $ l_j $ and $ \tilde{l}_j $,
$ j=1,2 $, have to be chosen such that the commutation relations
for the photon creation and annihilation operators are fulfilled.
These are $ [\hat{a}_j,\hat{a}_j^\dagger] = 1 $ for $ j=1,2 $ and
the remaining commutation relations among the operators $
\hat{a}_j $ and $ \hat{a}^\dagger_j $ are zero. Symbol $ \delta $
stands for the Dirac function. In case of damping in mode 1 and
the reservoir two-level atoms in the ground state, we have $ l_1 =
2\gamma_1 $ and $ \tilde{l}_1 = 0 $. On the other hand,
amplification in mode 2 and the reservoir two-level atoms in the
excited state gives $ l_2 = 0 $ and $ \tilde{l}_2 = 2|\gamma_2| $.
We note that, for standard PTSSs, amplification just compensates
for damping, i.e. $\gamma_1 = -\gamma_2 \equiv \gamma $. On the
other hand, we assume $\gamma_1 \equiv 2\gamma $ and $ \gamma_2
\equiv 0 $ for the analyzed passive PTSS. Similarly, we have
$\gamma_1 \equiv 0 $ and  $\gamma_2 \equiv -2\gamma $  for the
analyzed active PTSS. We also note that a more general PTSS
containing also the nonlinear Kerr and cross-Kerr terms was
analyzed in Refs.~\cite{PerinaJr2019x,PerinaJr2019y} using the
method of small operator corrections to mean values. We note that,
for the reservoir dynamics, we assume Markovian processes.
However, the analysis can also be extended to non-Markovian
dynamics. For example, a system coupled to a non-Markovian bath
can be modeled by introducing an ancilla that is linearly coupled
to the system and interacts with a standard Markovian reservoir.
Such an approach enables the analysis of non-Markovian quantum
exceptional points \cite{Lin2025}.

The Langevin-Heisenberg equations derived from the Hamiltonian $
\hat{H} $ in Eq.~(\ref{1}), with the help of the theory describing
the system interaction with the reservoir, can conveniently be
written in the following matrix form using the vectors $ \hat{\bf
A}^{\rm T} = (\hat{a}_1,\hat{a}_1^\dagger,
\hat{a}_2,\hat{a}_2^\dagger) $ and $ \hat{\bf L}^{\rm T} =
(\hat{l}_1,\hat{l}_1^\dagger, \hat{l}_2,\hat{l}_2^\dagger) $:
\begin{eqnarray} 
 \frac{d\hat{\bf A}(t)}{dt} &=& -i {\bf M} \hat{\bf A}(t)
  + \hat{\bf L}(t),
\label{3}  \\
 {\bf M} &=&   \left[ \begin{array}{cccc}
   -i\gamma_1 & 0 & \epsilon & \kappa \\
   0 & -i\gamma_1 & -\kappa & -\epsilon \\
   \epsilon & \kappa & -i\gamma_2 & 0 \\
   -\kappa & -\epsilon & 0 & -i\gamma_2 \end{array} \right].
\label{4}
\end{eqnarray}
In Eq.~(\ref{3}), we assume for simplicity real $ \epsilon $ and $
\kappa $.

We note that the model can readily be reformulated in terms of a
master equation in the Lindblad form. Since it is naturally
expressed in the coherent-state basis, this leads to a
Fokker-Planck equation for the corresponding quasi-probability
distribution. When assuming Gaussian states, the problem reduces
to solving a set of ordinary differential equations for the
first-order moments (mean amplitudes) and the second-order moments
[the statistical coefficients defined later in Eq.~(\ref{21})].

Introducing the evolution matrix $ {\bf P}(t,t')
$~\cite{PerinaJr2000},
\begin{equation} 
 {\bf P}(t,t') = \exp[ -i{\bf M} (t-t')],
\label{5}
\end{equation}
the solution of Eq.~(\ref{3}) is obtained as:
\begin{eqnarray}  
 \hat{\bf A}(t) &=& {\bf P}(t,0)\hat{\bf A}(0)
    + \hat{\bf F}(t),
\label{6}   \\
  \hat{\bf F}(t) &=& \int_{0}^t dt' {\bf P}(t,t')
  \hat{\bf L}(t'),
\label{7}
\end{eqnarray}
and $ \hat{\bf F}^{\rm T} \equiv
(\hat{f}_1,\hat{f}_1^\dagger,\hat{f}_2,\hat{f}_2^\dagger) $. The
properties of the Langevin fluctuating operator forces in
Eq.~(\ref{2}) result in the nonzero second-order correlation
functions of the forces $ \hat{\bf F} $ given
as~\cite{PerinaJr2000}:
\begin{eqnarray} 
  \langle \hat{\bf F}(t)^{} \hat{\bf F}^{\dagger {\rm T}}(t)\rangle &=& \int_{0}^t d\tilde{t}
  \int_{0}^t d\tilde{t}' {\bf P}(t,\tilde{t}) \langle \hat{\bf L}(\tilde{t})
  \hat{\bf L}^{\dagger {\rm T}}(\tilde{t}')\rangle\nonumber \\
 & & \times  {\bf P}^{\dagger {\bf T}}(t,\tilde{t}'),
\label{8}
\end{eqnarray}
where symbol $ T $ stands for the transposed matrix.

According to Eq.~(\ref{5}), the eigenvectors of the evolution
matrix $ {\bf P} $ coincide with those of the dynamical matrix $
{\bf M} $ and the corresponding eigenvalues $ \Lambda_{\bf P
}(t,t') $ are given as $ \exp[ -i\Lambda_{\bf M}(t-t')] $, where $
\Lambda_{\bf M} $ contains the eigenvalues of matrix $ {\bf M} $.
Diagonalization of the dynamical matrix $ {\bf M} $ leaves us
with the following eigenvectors and eigenvalues:
\begin{eqnarray} 
 {\bf M} &=& {\bf T} {\Lambda}_{\bf M} {\bf T}^{-1};
\label{9} \\
 & & \hspace{-10mm} {\Lambda}_{\bf M} =  -i\gamma_+ {\rm diag}(1,1,1,1) + \mu \;  {\rm diag}(1,1,-1,-1) ,
\label{10}  \\
 & & {\bf T} = ({\bf T}_1,{\bf T}_2,{\bf T}_3,{\bf T}_4),
\label{11} \\
 & & {\bf T}_{1,2}^{\rm T} = \frac{1}{2\sqrt{\epsilon}} \Bigl(
  \zeta^\pm, -\zeta^\mp, \pm \zeta^\pm \psi^+, \mp \zeta^\mp \psi^+
   \Bigr), \nonumber \\
 & & {\bf T}_{3,4}^{\rm T} = \frac{1}{2\sqrt{\epsilon}} \Bigl(
  \zeta^\pm, -\zeta^\mp, \mp \zeta^\pm \psi^-, \pm \zeta^\mp \psi^-
   \Bigr), \nonumber \\
 & & {\bf T}^{-1} = ({\bf T}^{-1}_1,{\bf T}^{-1}_2,{\bf T}^{-1}_3,{\bf T}^{-1}_4),\nonumber \\
 & & {\bf T}^{-1 {\rm T} }_{1,2} = \frac{\sqrt{\epsilon}}{2\sqrt{\mu}} \Bigl(
  \zeta^\pm \psi^-, - \zeta^\mp \psi^- ,
  \zeta^\pm \psi^+, - \zeta^\mp \psi^+   \Bigr), \nonumber \\
 & & {\bf T}^{-1 {\rm T}}_{3,4} = \frac{\sqrt{\epsilon}}{2\sqrt{\mu}} \left(
  \zeta^\pm, \zeta^\mp, -\zeta^\pm, -\zeta^\mp  \right), \nonumber
\end{eqnarray}
and $ \gamma_+ = (\gamma_1 + \gamma_2)/2 $, $ \gamma_- = (\gamma_1
- \gamma_2)/2 $,  $ \xi = \sqrt{\epsilon^2-\kappa^2} $, $
\zeta^\pm = \sqrt{\epsilon \pm \xi} $, $ \mu =
\sqrt{\epsilon^2-\kappa^2-\gamma_-^2} $, and $ \psi^\pm = (\mu \pm
i\gamma_-) /\xi $. We note that the structure of eigenvectors and
eigenvalues of the matrix $ {\bf M} $ closely resembles that
obtained for the special case $ \gamma_1 = -\gamma_2 $
($\mathcal{PT}$-symmetric case) analyzed in detail in
Ref.~\cite{PerinaJr2023} when studying the problem of
nonclassicality and entanglement losses in the long-time limit
caused by fluctuating forces and related to the properties of the
fluctuating forces.

According to Eq.~(\ref{10}) there exist two doubly degenerated
eigenvalues
\begin{equation}   
  \nu_{1,2} = -i(\gamma_1+\gamma_2)/2 \pm
\sqrt{ \epsilon^2 - \kappa^2 - (\gamma_1-\gamma_2)^2 /4}.
  \label{12}
\end{equation}
We have the real eigenvalues $ \nu_{1,2}^{\rm ad} $ for the
standard PTSS:
\begin{equation}  
 \nu_{1,2}^{\rm ad} = \sqrt{
 \epsilon^2 - \kappa^2 - \gamma^2 }.
\label{13}
\end{equation}
On the other hand, the eigenvalues $ \nu_j^{\rm dd} $ ($
\nu_j^{\rm aa} $) for the passive (active) PTSS additionally
contain a common damping (amplification) factor $ \gamma $ ($
-\gamma $):
\begin{eqnarray}  
 \nu_{1,2}^{\rm dd} &=& -i\gamma \pm \sqrt{
 \epsilon^2 - \kappa^2 - \gamma^2 },
\label{14} \\
 \nu_{1,2}^{\rm aa} &=& i\gamma \pm \sqrt{
 \epsilon^2 - \kappa^2 - \gamma^2 }
\label{15} .
\end{eqnarray}
Importantly, the eigenvectors of the matrix $ {\bf M} $
corresponding to the eigenvalues $ \nu_{1,2} $ as given in
Eq.~(\ref{11}) are identical for the standard, passive, and active
PTSSs. This means that, when we express the system evolution in
the basis of these eigenvectors, the dynamics in the three
discussed PTSSs differ just by common multiplicative functions
describing exponential damping in the passive PTSS and exponential
amplification in the active PTSS. This property provides a strong
foundation for a meaningful comparison of the statistical
characteristics of the three systems discussed below. It also
accounts for the fact that all three systems exhibit EPs at
identical locations in the parameter space. The existence and
degeneracies of these EPs---interpreted as Liouvillian EPs---were
examined in detail in Ref.~\cite{PerinaJr2022}.

The solution of the Heisenberg equations (\ref{3}) can then be
written in the following form that explicitly expresses the
symmetry contained in the above 4-dimensional matrix formulation:
\begin{equation}  
 \hat{\bf a}(t) = {\bf U}(t)\hat{\bf a}(0)
  + {\bf V}(t) \hat{\bf a}^\dagger(0) + \hat{\bf f}(t).
\label{16}
\end{equation}
In Eq.~(\ref{16}), we introduce $ \hat{\bf a}^{\rm T} \equiv
(\hat{a}_1,\hat{a}_2) $, $ U_{j,k}(t) = P_{2j-1,2k-1}(t,0) $, $
V_{jk}(t) = P_{2j-1,2k}(t,0) $, and $ \hat{f}_j(t) =
\hat{F}_{2j-1}(t) $, $ j,k=1,2 $. The matrices $ {\bf U} $ and $
{\bf V} $ attain the form:
\begin{eqnarray} 
 {\bf U}(t) &=&   \frac{1}{\mu} \left[ \begin{array}{cc}
  \mu c(t)-\gamma_- s(t) & -i\epsilon s(t) \\
  -i\epsilon s(t) & \mu c(t) +\gamma_- s(t) \end{array} \right] \exp(-\gamma_+ t),
  \nonumber \\
 {\bf V}(t) &=& - \frac{i\kappa s(t)}{\mu} \left[ \begin{array}{cc}
  0 & 1 \\ 1 & 0 \end{array} \right]  \exp(-\gamma_+ t),
\label{17}
\end{eqnarray}
where $ s(t) \equiv \sin(\mu t) $ and $ c(t) \equiv \cos(\mu t) $.

Incorporation of the solution into Eq.~(\ref{8}) for the matrix $
\langle \hat{\bf F}(t)^{} \hat{\bf F}^{\dagger {\bf T}}(t)\rangle
$ of correlation functions of the fluctuating forces results in
the following formulas for its elements:
\begin{eqnarray}   
 \langle \hat{f}_1^2(t)\rangle &=& (l_2 + \tilde{l}_2)
   \epsilon\kappa h(t) \theta , \nonumber \\
 \langle \hat{f}_1(t)\hat{f}_1^\dagger(t)\rangle &=& [l_1 h_-(t)  - (\epsilon^2l_2 + \kappa^2\tilde{l}_2)  h(t)]\theta , \nonumber \\
 \langle \hat{f}_1^\dagger(t)\hat{f}_1(t)\rangle &=& [\tilde{l}_1 h_-(t)
   - (\kappa^2 l_2 + \epsilon^2 \tilde{l}_2) h(t)]\theta , \nonumber \\
 \langle \hat{f}_2^2(t)\rangle &=& (l_1 + \tilde{l}_1)
   \epsilon\kappa h(t) \theta, \nonumber \\
 \langle \hat{f}_2(t)\hat{f}_2^\dagger(t)\rangle &=& [l_2 h_+(t)
  - (\epsilon^2l_1 + \kappa^2\tilde{l}_1)  h(t)] \theta , \nonumber \\
 \langle \hat{f}_2^\dagger(t)\hat{f}_2(t)\rangle &=& [\tilde{l}_2 h_-(t)
  - (\kappa^2 l_1 + \epsilon^2 \tilde{l}_1)  h(t)] \theta , \nonumber \\
 \langle \hat{f}_1(t)\hat{f}_2(t)\rangle &=& [l_1i\kappa d_-(t)
   + \tilde{l}_2i\kappa d_+(t)] \theta, \nonumber \\
 \langle \hat{f}_2(t)\hat{f}_1(t)\rangle &=& [\tilde{l}_1i\kappa d_-(t)
   + l_2i\kappa d_+(t)] \theta, \nonumber \\
 \langle \hat{f}_1^\dagger(t)\hat{f}_2(t)\rangle &=& [\tilde{l}_1i\epsilon d_-(t)
   - \tilde{l}_2i\epsilon d_+(t)] \theta, \nonumber \\
 \langle \hat{f}_2(t)\hat{f}_1^\dagger(t)\rangle &=& [l_1i\epsilon d_-(t)
   - l_2i\epsilon d_+(t)] \theta,
\label{18}
\end{eqnarray}
where $ \theta = 1 /(2\mu^2) $. We have introduced the following
functions in Eq.~(\ref{18}):
\begin{eqnarray}  
 f(t) &=& \frac{1-\exp(-2\gamma_+t) \exp(-2i\mu t) }{ 2(\gamma_+ +i\mu)},\nonumber\\
  g(t) &=& [1-\exp(-2\gamma_+t) ] / (2\gamma_+),\nonumber\\
  h(t) &=& \Re\{f(t)\} - g(t),\nonumber\\
  h_\pm(t) &=& \Re\{(\mu\pm i\gamma_-)^2f(t)\} + \xi^2 g(t),\nonumber\\
  d_\pm(t) &=& \Im\{(\mu\pm i\gamma_-)f(t)\} \mp \gamma_- g(t).
\label{19}
\end{eqnarray}
We note that we also have $ \langle \hat{\bf F}(t) \rangle =
\langle \hat{\bf F}^\dagger (t) \rangle = {\bf 0} $.

\section{Nonclassicality, entanglement, steering, and Bell nonlocality in
two-mode bosonic systems}

In the analysis of nonclassicality and quantum correlations, we
consider only the Gaussian states \cite{Perina1991,Adesso2007}
that, however, represent the most useful states both in the
analysis of fundamental physical experiments and applications.
Moreover and most importantly, the linear Heisenberg-Langevin
equations in Eq.~(\ref{3}) describe the state evolution inside
this group of states. They are conveniently described by their
normal characteristic function $ C_{\mathcal N} $ written in the
general form as~\cite{Perina1991,Adesso2007}:
\begin{eqnarray} 
 C_{\cal N}(\mu_1,\mu_2,t) &=& \exp\Biggl\{ \sum_{j=1,2} \Bigl[
   \left( \alpha_j^*(t)\mu_j - {\rm c.c.}\right)
   \nonumber\\
 & & \hspace{-15mm} \mbox{} -B_j(t)|\mu_j|^2 + \frac{C_j(t)\mu_j^{2*} + {\rm c.c.}}{2} \Biggr]
  \nonumber \\
 & & \hspace{-15mm} \mbox{} + \left[ D(t)\mu_1^*\mu_2^* + \bar{D}(t)\mu_1\mu_2^* + {\rm
   c.c.}\right] \Bigr\},
\label{20}
\end{eqnarray}
and c.c. replaces the complex conjugated term.

The parameters $ B_j $, $ C_j $, $ D $, and $ \bar{D} $ that,
together with the mode complex amplitudes, identify the state are
obtained according to the formulas valid for the initial coherent
states with amplitudes $ \alpha_1(0) $ and $ \alpha_2(0) $:
\begin{eqnarray}  
 B_j(t) &\equiv& \langle\delta\hat{a}_j^\dagger(t)\delta\hat{a}_j(t)\rangle
  = \sum_{l=1,2} \left[ |V_{jl}(t)|^2 +
  \langle\hat{f}_j^\dagger(t)\hat{f}_j(t)\rangle \right], \nonumber \\
 C_j(t) &\equiv& \langle [\delta\hat{a}_j^2(t)]^2\rangle
  = \sum_{l=1,2} \left[ U_{jl}(t) V_{jl}(t) +
  \langle\hat{f}_j^2(t)\rangle \right], \nonumber \\
 D(t) &\equiv& \langle\delta\hat{a}_1(t)\delta\hat{a}_2(t)\rangle
   \nonumber \\
 &=& \sum_{l=1,2} \left[ U_{1l}(t)V_{2l}(t) +
  \langle\hat{f}_1(t)\hat{f}_2(t)\rangle \right], \nonumber \\
 \bar{D}(t) &\equiv& -\langle\delta\hat{a}_1^\dagger(t)\delta\hat{a}_2(t)\rangle
  \nonumber \\
 &=& -\sum_{l=1,2} \left[V_{1l}^*(t)V_{2l}(t)  +
   \langle\hat{f}_1^\dagger(t)\hat{f}_2(t)\rangle \right],
\label{21}
\end{eqnarray}
where $ \delta \hat{a}_j = \hat{a}_j - \langle \hat{a}_j \rangle $
for $ j=1,2 $. Substituting Eq.~(\ref{17}) into Eq.~(\ref{21}), we
arrive at the formulas appropriate for our model:
\begin{eqnarray}   
 B_j(t) &=& (\kappa/\mu)^2 \tilde{s}(t) + \langle f_j^\dagger(t)f_j(t)\rangle,\nonumber \\
 C_j(t) &=& -(\epsilon\kappa/\mu^2) \tilde{s}(t) + \langle f_j^2(t)\rangle, \hspace{5mm} j=1,2,\nonumber \\
 D(t) &=& -i(\kappa/\mu) \tilde{c}(t) + i(\kappa\gamma_+/\mu^2) \tilde{s}(t) + \langle f_1(t)f_2(t)\rangle,\nonumber \\
 \bar{D}(t) &=& -\langle f_1^\dagger(t)f_2(t)\rangle,
\label{22}
\end{eqnarray}
where $ \tilde{s}(t) = \sin^2(\mu t) \exp(-2\gamma_+ t) $, $
\tilde{c}(t) = \sin(\mu t)\cos(\mu t) \exp(-2\gamma_+ t) $, and
the correlation functions of the fluctuating forces are given in
Eq.~(\ref{18}).

At an EP, we have $ \mu = 0 $ and the formulas (\ref{22}) for the
statistical parameters considerably simplify ($ \mu \rightarrow 0
$):
\begin{eqnarray} 
 B_j^{\rm EP}(t) &=& \kappa^2 t^2 \exp(-2\gamma_+ t) + \tilde{l}_j
   g(t)/2 - (\kappa^2 l_{3-j} \nonumber \\
  & &   + \epsilon^2 \tilde{l}_{3-j}) \tilde{h}_0(t)/2, \nonumber \\
 C_j^{\rm EP}(t) &=& -\epsilon\kappa t^2 \exp(-2\gamma_+ t) + (l_{3-j} + \tilde{l}_{3-j}) \epsilon\kappa \tilde{h}_0(t)/2, \nonumber \\
 & & \hspace{2cm} j=1,2, \nonumber \\
 D^{\rm EP}(t) &=& -i \kappa (t - \gamma_+ t^2) \exp(-2\gamma_+ t) + (l_1 +
 \tilde{l}_2)i\kappa \nonumber \\
  & & \times [ t\exp(-2\gamma_+ t) - g(t) ]/(2\gamma_+) - (l_1 - \tilde{l}_2)
  \nonumber \\
 & & \times i\kappa \gamma_-\tilde{h}_0(t)/2, \nonumber \\
 \bar{D}^{\rm EP}(t) &=& (\tilde{l}_2 - \tilde{l}_1)i\epsilon [ t\exp(-2\gamma_+ t) - g(t) ]/(2\gamma_+)
  \nonumber \\
  & &  + (\tilde{l}_1 + \tilde{l}_2)  i\epsilon
  \gamma_-\tilde{h}_0(t)/2,
\label{23}
\end{eqnarray}
where $ \tilde{h}_0(t) \equiv [ (t + \gamma_+ t^2) \exp(-2\gamma_+
t) -g(t)] / \gamma_+^2 $. We can see that the original oscillatory
behavior of the standard PTSS is replaced by the polynomial one at
the EP. For the passive (active) PTSS additional exponential
damping (amplification) occurs.

\subsection{Nonclassicality}

Coefficients of the quadratic terms in the argument of exponential
function in Eq.~(\ref{20}) can be arranged into the matrix $ {\bf
K}_{s} $ using the vector $ (\beta_1,\beta_1^*,\beta_2,\beta_2^*)
$. The matrix $ {\bf K}_{s} $ describes the characteristic
function $ C_{s} $ written in the general $ s $ ordering of the
field operators \cite{Perina1991}:
\begin{equation}   
 {\bf K}_{C_s}(s) = \frac{1}{2} \left[ \begin{array}{cccc}
   -B_{1,s}(s) & C_1^* & \bar{D}^* & D \\
   C_1 &  -B_{1,s}(s) & D^* & \bar{D} \\
   \bar{D} & D & -B_{2,s}(s) & C_2^* \\
   D^* & \bar{D}^* & C_2 &  -B_{2,s}(s) \end{array} \right],
\label{24}
\end{equation}
and $ B_{j,s}(s) = (1-s)/2 + B_j $ for $ j=1,2 $. Eigenvalues of
the matrix $ {\bf K}_{C_s}(s) $, that depend on the ordering
parameter $ s $, bear the information about the state
nonclassicality \cite{Glauber1963,Mandel1995}. Detailed analysis
reveals that the Lee nonclassicality depth $ \tau $
\cite{Lee1991,PerinaJr2023} of the state is equal to the greatest
positive eigenvalue of the matrix $ {\bf K}_{C_s}(s=1) $ written
for the normal field-operator ordering. Applying this procedure to
the individual modes, we immediately arrive at the formula for the
local nonclassicality depths $ \tau_j $ for $ j=1,2 $:
\begin{equation}   
 \tau_j = {\rm max}\{ 0,|C_j| - B_j\}.
\label{25}
\end{equation}

\subsection{Steering}

Quantum correlations of the Gaussian states are described by their
coherence matrix $ \sigma $ defined for the vector $
(\hat{q}_1,\hat{p}_1,\hat{q}_2,\hat{p}_2) $ \cite{Adesso2007}:
\begin{eqnarray}  
 {\bf \sigma} &=& \left[ \begin{array}{cc}
  {\bf \sigma}_1 &  {\bf \sigma}_{12} \\
  \left[{\bf \sigma}_{12}\right]^{T} &  {\bf \sigma}_2 \end{array}
  \right] ,
\label{26} \\
  {\bf \sigma}_j &=& \left[ \begin{array}{cc}
   1+2B_j+2\Re\{C_j\} & 2\Im\{C_j\} \\
   2\Im\{C_j\} &   1+2B_j-2\Re\{C_j\} \end{array}
  \right] , \nonumber \\
  {\bf \sigma}_{12} &=& 2\left[ \begin{array}{cc}
   \Re\{D-\bar{D}\} & \Im\{D - \bar{D}\} \\
   \Im\{D+\bar{D}\} & - \Re\{D+\bar{D}\} \end{array}
  \right] ,
\label{27}
\end{eqnarray}
where $ \hat{q}_j = (\hat{a}_j + \hat{a}_j^\dagger)/2 $, $
\hat{p}_j = (\hat{a}_j - \hat{a}_j^\dagger)/(2i) $, $ j=1,2 $. The
steering of mode $ (3-j) $ by mode $ j $ for $ j=1,2 $ is then
expressed using the formula \cite{Cavalcanti2009}:
\begin{equation}  
 S_{j\rightarrow 3-j} = \max\{ 0, \det\{\sigma_j\} / \det\{\sigma\}
  \}/2.
\label{28}
\end{equation}

\subsection{Entanglement}

The (logarithmic) negativity $ E_N $ \cite{Hill1997,Horodecki2009}
as a commonly accepted measure of entanglement is inferred from
the coherence matrix $ \sigma^{PT} $ defined for the vector $
(\hat{q}_1,\hat{p}_1,\hat{q}_2,-\hat{p}_2) $, i.e. for the
partially transposed state of mode 2,
\cite{Peres1996,Horodecki1997,Simon2000}:
\begin{eqnarray}  
 {\bf \sigma}^{PT} &=& \left[ \begin{array}{cc}
  {\bf \sigma}_1 &  {\bf \sigma}_{12}^{PT} \\
  \left[{\bf \sigma}_{12}^{PT}\right]^{T} &  {\bf \sigma}_2^{PT} \end{array}
  \right] ,
\label{29} \\
  {\bf \sigma}_2^{PT} &=& \left[ \begin{array}{cc}
   1+2B_2+2\Re\{C_2\} & -2\Im\{C_2\} \\
   -2\Im\{C_2\} &   1+2B_2-2\Re\{C_2\} \end{array}
  \right] , \nonumber \\
  {\bf \sigma}_{12}^{PT} &=& 2\left[ \begin{array}{cc}
   \Re\{D-\bar{D}\} & \Im\{-D+\bar{D}\} \\
   \Im\{D+\bar{D}\} & \Re\{D+\bar{D}\} \end{array}
  \right] .
\label{30}
\end{eqnarray}
The symplectic eigenvalue $ \nu_{-} $ determined with the help of
the invariants $ \Delta $ and $ \delta $ \cite{Adesso2007},
\begin{eqnarray}  
 \nu_{-} &=& \sqrt{ \frac{\delta}{2} - \sqrt{ \frac{\delta^2}{4}
   - \Delta }},
 \label{31} \\
 \Delta &=& {\rm det}\{ {\bf \sigma}^{PT} \} , \nonumber \\
 \delta &=& {\rm det}\{ {\bf \sigma}_1\} + {\rm det}\{ {\bf
\sigma}_2^{PT}\} + 2{\rm det}\{ {\bf \sigma}_{12}^{PT}\} ,
\nonumber
\end{eqnarray}
then gives the negativity:
\begin{equation}  
 E_N = {\rm max}\{0,-\ln(\nu_-)\}.
\label{32}
\end{equation}

\subsection{Bell nonlocality}

The strongest quantum correlations, that imply the Bell
nonlocality, are quantified by the Bell parameter $ B_{\rm Bell} $
\cite{Bell1964} that is in our case a specific linear combination
of the mean values of suitably displaced parity operators $
\hat{\Pi}(\beta_1,\beta_2) $. Introducing two sets of
displacements $ (\beta_1,\beta_2) $ and $ (\beta'_1,\beta'_2) $
the Bell parameter $ B_{\rm Bell} $ is determined along the
formula:
\begin{eqnarray}    
 B_{\rm Bell}(\beta_1,\beta_2;\beta'_1,\beta'_2) &=&
  \langle\hat{\Pi}(\beta_1,\beta_2)\rangle +
  \langle\hat{\Pi}(\beta'_1,\beta_2)\rangle \nonumber \\
  & & \hspace{-5mm} + \langle\hat{\Pi}(\beta_1,\beta'_2)\rangle -
  \langle\hat{\Pi}(\beta'_1,\beta'_2)\rangle .
\label{33}
\end{eqnarray}
If $ |B_{\rm Bell}| > 2 $ for any suitable choice of the
displacements $ (\beta_1,\beta_2) $ and $ (\beta'_1,\beta'_2) $,
the state exhibits the Bell nonlocality manifested by the
violation of the Bell inequalities \cite{Bell1964}. According to
Refs.~\cite{Banaszek1998,Banaszek1999}, the mean value of a
displaced parity operator is directly obtained from the Wigner
function $ \Phi_{s=0} $ using the formula:
\begin{equation}   
 \langle\hat{\Pi}(\beta_1,\beta_2)\rangle = \frac{\pi^2}{4}
   \Phi_{s=0}(\beta_1,\beta_2) .
\label{34}
\end{equation}
To reveal the Wigner function $ \Phi_{s=0} $, we first have to
rewrite the matrix $ {\bf K}_{C_s}(s=0) $ into that written for
the vector $ \bm{\alpha}^{{\rm real},T} \equiv
(\Re\{\alpha_1\},\Im\{\alpha_1\},\Re\{\alpha_2\},\Im\{\alpha_2\})
$:
\begin{eqnarray}   
 {\bf K}_{C_s}^{\rm real} &=& \left[ \begin{array}{cc}
   -B_{1,s}(0)+\Re\{C_1\} & \Im\{C_1\} \\
   \Im\{C_1\} & -B_{1,s}(0)-\Re\{C_1\} \\
   \Re\{D+\bar{D}\} & \Im\{D-\bar{D}\} \\
   \Im\{D+\bar{D}\} & \Re\{-D+\bar{D}\} \end{array} \right. \nonumber \\
   & & \hspace{-5mm} \left. \begin{array}{cc}
    \Re\{D+\bar{D}\} & \Im\{D+\bar{D}\} \\
    \Im\{D-\bar{D}\} & \Re\{-D+\bar{D}\} \\
     -B_{2,s}(0)+\Re\{C_2\} & \Im\{C_2\} \\
   \Im\{C_2\} & -B_{2,s}(0)-\Re\{C_2\}
   \end{array} \right].
\label{35}
\end{eqnarray}
Forming the vector $ \bm{\alpha}^{{\rm real},T} \equiv
(\Im\{\alpha_1\},\Re\{\alpha_1\},\Im\{\alpha_2\},\Re\{\alpha_2\})
$) from the arguments $ \alpha_1 $ and $ \alpha_2 $ of the Wigner
function $ \Phi_{s=0}(\alpha_1,\alpha_2) $ the 4-dimensional
Fourier transform of the characteristic function $
C_{s=0}(\bm{\alpha}^{{\rm real},T}) $ related to the symmetric
ordering of field operators leaves the Wigner function in the
form:
\begin{equation}    
 \Phi_{s=0}(\alpha_1,\alpha_2) = \frac{ \exp\left[ \bm{\alpha}^{{\rm real},T} {\bf K}_{C_s}^{{\rm   real}, -1}
  \bm{\alpha}^{\rm real} \right]  }{ \pi^2 \sqrt{ \det\{ {\bf K}_{C_s}^{\rm real}\} } }.
\label{36}
\end{equation}
In Eq.~(\ref{36}), we use the inverse to the matrix $ {\bf
K}_{C_s}^{{\rm real}} $ and its determinant.

The Bell parameter $ B_{\rm Bell} $ depends on two sets of
displacements that have to be suitably chosen to reveal the
violation of the Bell inequalities. In the numerical analysis,
inspired by Refs.~\cite{Banaszek1998,Olivares2004,Thapliyal2021},
we set $ (\beta_1,\beta_2) = (0,0)$ and systematically scan the
remaining complex displacements $ (\beta'_1,\beta'_2) $ (expressed
in radial coordinates) such that $ |\beta'_j| \le 2\sqrt{
B_{j,s}(0) } $ for $ j=1,2 $.

\section{Role of $\mathcal{PT}$-symmetry in nonclassical-state generation}

Before diving into a detailed discussion of the system's behavior,
it is important to note that its dynamics---particularly regarding
damping and amplification---are shaped by two competing effects.
The first involves the influence of damping or amplification on
the coherent component of the system's evolution. In this context,
amplification is generally considered beneficial compared to
damping, as it increases mode amplitudes. This increase, in turn,
enhances the system's effective  {physical} nonlinearity, defined
as the product of the nonlinear coupling constant and the mode
amplitudes.

The second effect stems from random fluctuating forces (i.e.,
quantum noise) that tend to degrade nonclassical features and
quantum correlations by disrupting phase coherence within the
system. Notably, the noise accompanying amplification is generally
stronger than that associated with damping, due in part to
spontaneous photon emission from reservoir atoms in excited
states.

In detail, mutual balance between damping and amplification in
standard PTSSs gives their specific dynamical behavior. In our
case, $\mathcal{PT}$-symmetry is reached assuming $\gamma_1 =
\gamma $, $ \gamma_2 = -\gamma $, $ l_1 = 2\gamma $, $ \tilde{l}_2
= 2\gamma $, $ \tilde{l}_1=l_2 = 0 $. In specific cases, at EPs,
spectral degeneracies of the dynamical matrix $ {\bf M} $ occur
being accompanied by the corresponding eigenvector degeneracies.
This means that the system evolution considerably simplifies and
only a single eigenfrequency is sufficient to describe the
evolution. Following Eq.~(\ref{10}), or Eq.~(\ref{13}), such
situation occurs provided that $ \mu \equiv \sqrt{ \epsilon^2 -
\kappa^2 - \gamma^2 } = 0 $. For the Hamiltonian $ \hat{H} $ in
Eq.~(\ref{1}) and assuming $ \gamma_1 = -\gamma_2 = \gamma $, EPs
occur for
\begin{equation}  
 \frac{\kappa^2}{\epsilon^2} + \frac{\gamma^2}{\epsilon^2} =1 .
\label{37}
\end{equation}
In the analyzed system, this specific dynamics influences the
ability to generate nonclassical states of different kinds. The
nonclassicality of a generated state reflects either local
nonclassicalities of the constituting modes 1 and 2 or quantum
correlations between these modes. Whereas we quantify below the
nonclassicalities by the corresponding Lee nonclassicality depths
$ \tau $, $ \tau_1 $ and $ \tau_2 $, quantum correlations with the
increasing quantumness are in turn quantified by the negativity $
E_N $ (entanglement), steering parameters $ S_{1\rightarrow 2} $
and $ S_{2\rightarrow 1} $ (steering), and the Bell parameter
(Bell nonlocality).

We examine the system behavior by assuming the modes initially in
their vacuum states (arbitrary initial coherent states in both
modes can be considered as well) and follow their temporal
evolution. To assess the behavior of the investigated quantities,
we determine their maximal values along the time $ t $ axis:
\begin{eqnarray} 
 &\tau = \max_{t\epsilon}\{\tau(t\epsilon)\}, \quad E_N = \max_{t\epsilon}\{E_N(t\epsilon)\},& \nonumber \\
 &\tau_j = \max_{t\epsilon}\{\tau_j(t\epsilon)\}, \quad S_{j\rightarrow 3-j} = \max_{t\epsilon}\{
   S_{j\rightarrow 3-j}(t\epsilon)\},& \nonumber \\
 & B_{\rm Bell} = \max_{t\epsilon}\{B_{\rm Bell}(t\epsilon)\} ,
\label{38}
\end{eqnarray}
where $ j=1,2 $ and compare these maximal values for the whole
parameter space of the investigated system. We note that, due to
linearity of the corresponding Heisenberg equations, the parameter
space is effectively 2-dimensional and is spanned by the variables
$ \kappa/\epsilon $ and $ \gamma/\epsilon $.

To reveal the role of balance between damping and amplification in
the standard PTSS, we compare its behavior with two specific cases
(systems) in which only damping ($\gamma_1 =\gamma $, $ \gamma_2 =
0 $, $ l_1 = 2\gamma $, $ \tilde{l}_1=l_2=\tilde{l}_2 = 0 $) and
only amplification ($\gamma_1 = 0 $, $ \gamma_2 = -\gamma $, $
\tilde{l}_2 = 2\gamma $, $ l_1=\tilde{l}_1=l_2= 0 $) influence the
system dynamics. Both systems for comparison, owing to their
 {physical} nonlinearities, preserve the ability to generate
the nonclassical states. Mutual comparison of the maximal values
of the quantities characterising both nonclassicality and
different types of quantum correlations then sheds light on how
beneficial the balance between damping and amplification in
standard PTSSs is when generating the nonclassical states. We note
that in the systems for comparison different conditions for EPs
occur:
\begin{equation}  
 \frac{\kappa^2}{\epsilon^2} + \frac{\gamma^2}{4\epsilon^2} =1 .
\label{39}
\end{equation}

The change of the system dynamics caused by the absence of
amplification even results in the observation of nonclassical
properties of the modes in the asymptotic limit $ t\rightarrow
\infty $ [$ t\epsilon\rightarrow \infty $]. In this case, only the
following coefficients from Eq.~(\ref{22}) attain asymptotically
nonzero values:
\begin{equation} 
  B_2(\infty) = \frac{\kappa^2}{\epsilon^2 -\kappa^2}, \hspace{2mm}
 C_2(\infty) = - \frac{\epsilon\kappa}{\epsilon^2 -\kappa^2}.
\label{40}
\end{equation}
They imply the nonclassicality in mode 2 quantified by the
nonclassicality depth $ \tau_2 $:
\begin{equation}  
  \tau_2(\infty) = \frac{\kappa}{\epsilon+\kappa} .
\label{41}
\end{equation}

\subsection{Nonclassicality}

\begin{figure*}[t]  
 \centerline{ \includegraphics[width=0.98\hsize]{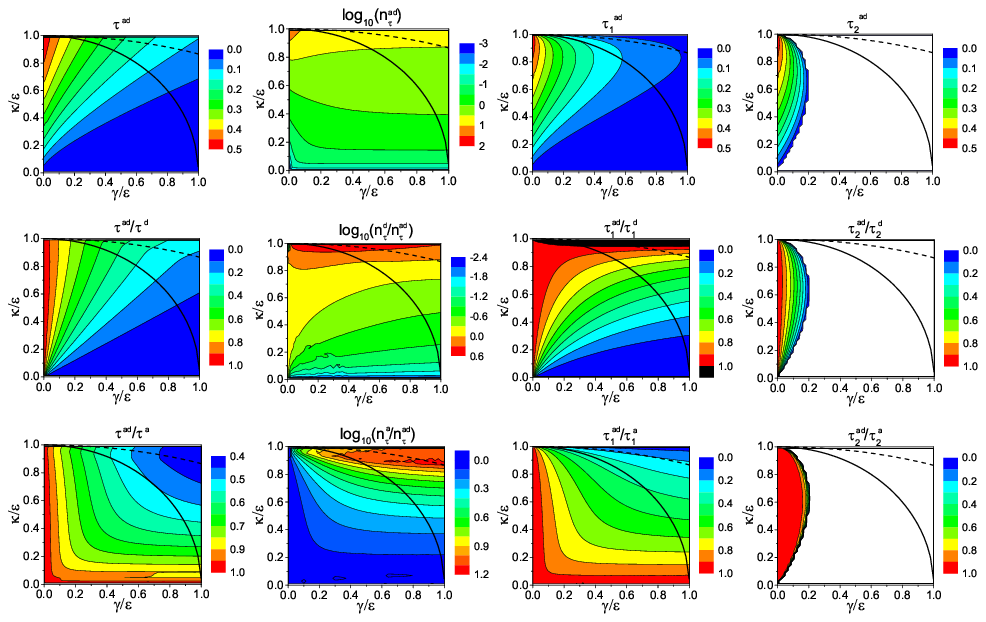} }
 \vspace{2mm}
 \centerline{ \small (a) \hspace{.22\hsize} (b) \hspace{0.22\hsize} (c) \hspace{0.22\hsize} (d)}
 \caption{(a) Nonclassicality depth $ \tau^{\rm ad} $ and (b) the corresponding mean photon
   number $ n_{\tau}^{\rm ad} $, (c) [(d)] local nonclassicality depth $ \tau_{1}^{\rm ad} $ [$ \tau_{2}^{\rm ad} $] of mode 1 [2]
   of standard PTSS as they depend on model parameters
   $ \gamma/\epsilon $ and $ \kappa/\epsilon $. The values of the
   drawn parameters are compared with those originating in the
   model with considered only damping (superscript d) and only
   amplification (superscript a).
   In white areas, $ \tau_{2}^{\rm ad} = 0 $.
   Solid [dashed] black curves identify positions of EPs
   in PTSS [systems with only damping and only amplification].
   The superscript notation is explained in Fig.~\ref{fig1}.}
 \label{fig2}
\end{figure*}

As documented in Figs.~\ref{fig2}(a-d, first row), the analyzed
standard PTSS allows for the generation of highly nonclassical
states with $ \tau, \tau_1, \tau_2 \rightarrow 1/2 $ for small $
\gamma/\epsilon $ and $ \kappa/\epsilon $ close to 1, i.e. when
the system damping and amplification are small. We note that 1/2
gives the greatest value of nonclassicality depth attained by a
Gaussian state.
As the nonlinear coupling constant $ \kappa $ represents the
source of nonclassicality, the greater the value of $ \kappa $ is
the better the ability of the system to generate nonclassical
states is. Despite the balance between damping and amplification,
the greater the damping and amplification are the worse the system
ability to provide nonclassical states is. This is because
stronger damping and amplification are accompanied by more intense
fluctuating forces. This relationship is quantified by the
fluctuation-dissipation and fluctuation-amplification theorems
\cite{Meystre2007}. These fluctuating forces then weaken the
system ability to generate nonclassical states. We can see in
Fig.~\ref{fig2}(b) that we also reach greatest field intensities $
n^{\rm ad} $ in the area of parameters optimal for
nonclassical-state generation.

The comparison of maximal values of the nonclassicality depths $
\tau $ and $ \tau_1 $ with those characterizing the systems with
only damping [see Figs.~\ref{fig2}(a,c, second row)] and only
amplification [see Figs.~\ref{fig2}(a,c, third row)] leads us to
the conclusion that both systems for comparison provide greater
maximal values than those of the corresponding standard PTSS for
most system parameters. The only exception is the narrow region in
the graph of the ratio $\tau_1^{\rm ad}/\tau_1^{\rm d}$ of
nonclassicality depths [see Fig.~\ref{fig2}(c), second row] for
values of $ \kappa/\epsilon $ close to 1. In this region, a large
nonlinear coupling constant $ \kappa $, combined with the high
amplitudes of the amplified mode 2 in the standard PTSS, leads to
enhanced effective  {physical} nonlinearity. This, in turn,
results in stronger nonclassicality compared to the case without
the amplified mode 2. Notably, this region naturally occurs near
the curve of EPs and within the domain of exponential mode
amplitude growth, in contrast to the periodic amplitude behavior
observed in the $\mathcal{PT}$-symmetric region.

On the other hand, for fixed model parameters, the nonclassicality
depth $\tau_2$ of mode~2 is highest for the system with only
damping [see Fig.~2(d), second row] and lowest for the system with
only amplification [see Fig.~2(d), third row]. Moreover, in the
area of parameters optimal for nonclassical-state generation, both
systems for comparison give more intense (and more nonclassical)
nonclassical states.

\subsection{Entanglement and steering}

\begin{figure*}[t]  
 \centerline{\includegraphics[width=0.98\hsize]{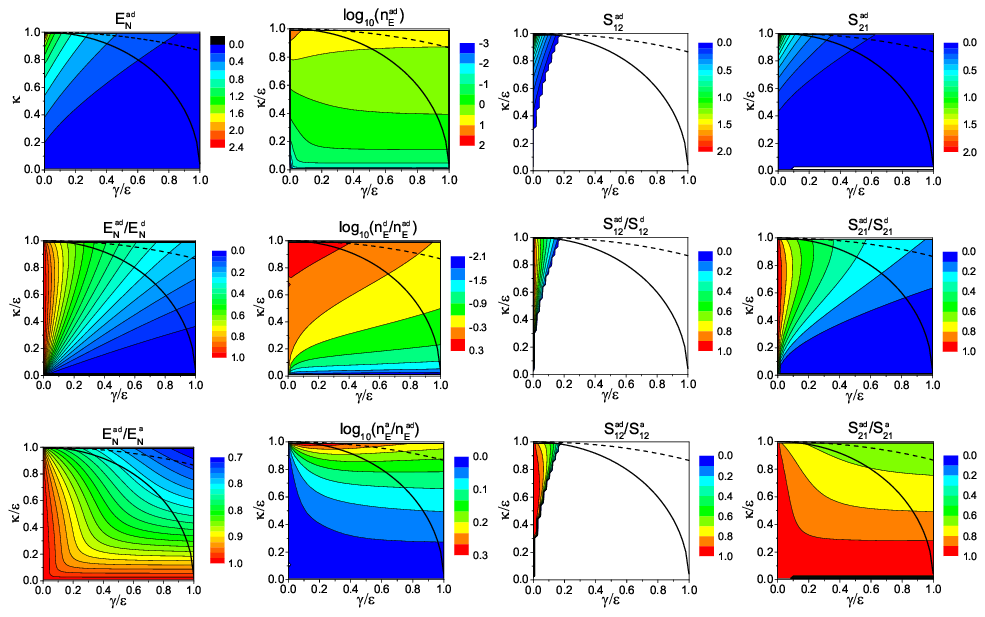}}
 \vspace{2mm}
 \centerline{ \small (a) \hspace{.22\hsize} (b) \hspace{0.22\hsize} (c) \hspace{.22\hsize} (d)}
 \caption{(a) Negativity $ E_N^{\rm ad} $ and (b) the corresponding mean photon
   number $ n_{E}^{\rm ad} $, (c) [(d)] steering parameter $ S_{1\rightarrow
   2}^{\rm ad} $ [$ S_{2\rightarrow 1}^{\rm ad} $]
   of standard PTSS as they depend on model parameters
   $ \gamma/\epsilon $ and $ \kappa/\epsilon $. The values of the
   drawn parameters are compared with those originating in the
   model with considered only damping (superscript d) and only
   amplification (superscript a). In white areas, $ S_{1\rightarrow
   2}^{\rm ad} = 0 $. Solid [dashed] black curves identify positions of EPs
   in PTSS [systems with damping and amplification].
   The superscript notation is explained in Fig.~\ref{fig1}.}
 \label{fig3}
\end{figure*}

The analysis of entanglement quantified by the negativity $ E_N $
and steering described by the steering parameters $
S_{1\rightarrow 2} $ and $ S_{2\rightarrow 1} $ provides us the
graphs in Figs.~\ref{fig3}(a-d, first row) for the standard PTSS
and the graphs in Figs.~\ref{fig3}(a-d, second row) when the
system with only damping is considered and the graphs in
Figs.~\ref{fig3}(a-d, third row) when the system with only
amplification is addressed.
The conclusions drawn from these graphs are similar to the above
ones for the nonclassicality depth $ \tau $: Both systems for
comparison allow for greater values of the negativity $ E_N $ and
the steering parameters $ S_{1\rightarrow 2} $ and $
S_{2\rightarrow 1} $ than the standard PTSS. We note that, whereas
greater nonlinearity constant $ \kappa/\epsilon $ and small
damping and amplification constants are required to allow steering
of the amplified mode 2 by the damped mode 1, the amplified mode 2
steers the damped mode 1 for any value of the system parameters.

\subsection{Bell nonlocality}

\begin{figure*}[ht]  
 \centerline{ \includegraphics[width=0.8\hsize]{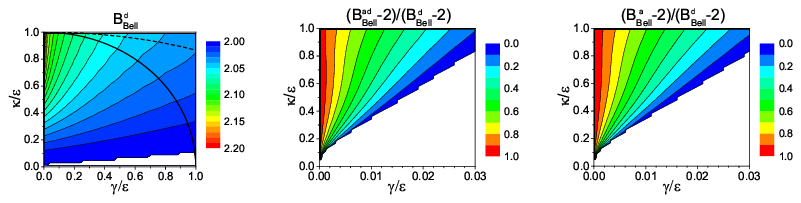}}
 \vspace{2mm}
 \centerline{ \small (a) \hspace{.22\hsize} (b) \hspace{0.22\hsize} (c)}
 \vspace{2mm}
 \centerline{ \includegraphics[width=0.6\hsize]{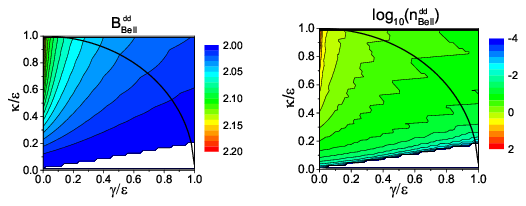} }
 \vspace{2mm}
 \centerline{ \small (d) \hspace{.22\hsize} (e)}
 \caption{(a) Bell parameter $ B_{\rm Bell}^{\rm d} $, (b) [(c)]
  ratio $ (B_{\rm Bell}^{\rm ad}-2)/(B_{\rm Bell}^{\rm d}-2) $
  [$ (B_{\rm Bell}^{\rm a}-2)/(B_{\rm Bell}^{\rm d}-2) $] of Bell
  parameters, (d) Bell parameter $ B_{\rm Bell}^{\rm dd} $ and
  the corresponding mean photon number $ n_{\rm Bell}^{\rm dd} $
  as they depend on model parameters  $ \gamma/\epsilon $ and $ \kappa/\epsilon $.
  In white areas, the Bell inequalities are not violated ($ B_{\rm Bell} = 2 $).
   Solid [dashed] black curves identify positions of EPs
   in PTSS as well as systems with doubled damping and
   amplification [systems with damping and amplification].
   The superscript notation is explained in Fig.~\ref{fig1}.}
 \label{fig4}
\end{figure*}

The advantage of the system with only damping in
nonclassicality-state generation over the other two investigated
systems manifests dramatically when generating the states that
exhibit the Bell nonlocality. Only this system allows to violate
the Bell inequalities in the wide area of the system parameters:
Only the small nonlinearity constant $ \kappa/\epsilon $ and
greater damping and amplification constants $ \gamma/\epsilon $
prevent from the violation of the Bell inequalities [see
Fig.~\ref{fig4}(a)]. Contrary to this, only very small values of
damping and amplification constants $ \gamma/\epsilon $ are
compatible with the states violating the Bell inequalities in the
standard PTSS and its variant with only amplification [see
Figs.~\ref{fig4}(b,c), $ \gamma/\epsilon <\approx 0.03 $]. Even
under these conditions the attained values of the Bell parameters
$ B_{\rm Bell}^{\rm ad} $ and $ B_{\rm Bell}^{\rm a} $ are smaller
than the parameters $ B_{\rm Bell}^{\rm d} $ belonging to the
system with only damping.

In summary, the coexistence of damping and amplification under
balanced conditions in the standard PTSS offers a clear advantage
only when the nonlinear coupling constant $ \kappa $ is large. In
this case, the increased amplitudes of the amplified mode 2
enhance the system's effective  {physical} nonlinearity, leading
to higher nonclassicality depths $ \tau_1 $ for the damped mode 1.
However, from the perspective of other quantumness
quantifiers---such as the global nonclassicality depth,
negativity, steering parameters, and the Bell nonlocality
parameter---this balance provides no significant benefit.

\section{The role of quantum fluctuations in nonclassical-state generation}

Parallel eigenvalue analysis of the dynamical matrices $ {\bf M} $
of the standard PTSS [damping constant $ \gamma_1 =\gamma $ and
amplification constant $ \gamma_2 = -\gamma $, for the
eigenvalues, see Eq.~(\ref{13})], passive PTSS [damping constant $
\gamma_1 = 2\gamma $ and no amplification $ \gamma_2 = 0 $, $ l_1
= 4\gamma $, $ \tilde{l}_1=l_2=\tilde{l}_2 = 0 $, see
Eq.~(\ref{14}], and active PTSS [no damping $ \gamma_1 = 0 $ and
amplification constant $ \gamma_2 = -2\gamma $, $ \tilde{l}_2 =
4\gamma $, $ l_1 = \tilde{l}_1=l_2= 0 $, see Eq.~(\ref{15}]
reveals striking similarity.

Their eigenfrequencies $ \Lambda_{\bf M} $ differ just by their
imaginary parts common to all eigenvalues: Whereas the imaginary
part of $ \Lambda_{\bf M} $ is zero for the standard PTSS, it
gives the average damping constant $ \gamma $ for the passive
system and the average amplification constant $ -\gamma $ for the
active system. Also, as already discussed in Sec.~III, the
corresponding eigenvectors are the same. We note that detailed
analysis of such behavior is given in Ref.~\cite{Chimczak2023}.
This similarity means that, for the same initial conditions for
modes 1 and 2, the evolution of operator amplitudes of the passive
[active] system differs from that of the standard PTSS just by the
multiplicative function $ \exp(-\gamma t ) $ [$ \exp(\gamma t )
$]. When the coefficients of the normal characteristic function $
C_{\cal N} $ given in Eq.~(\ref{22}) are considered the
multiplicative factors are $ \exp(-2\gamma t ) $ [$ \exp(2\gamma t
) $]. It is not only this coherent dynamics that causes different
evolution of the three PTSS. The dynamics of these systems differ
also because of different properties of their fluctuating operator
forces $ \hat{l}_j $ and $ \hat{l}_j^\dagger $, $ j=1,2 $,
prescribed to the modes in Eq.~(\ref{2}). Especially, the
properties of fluctuating forces assigned to the modes with
amplification have more detrimental influence to the
nonclassical-state generation than those belonging to the modes
with damping because of spontaneous emission in the reservoir
modes \cite{Scheel2018}. We note that, when a mode is neither
damped nor amplified no fluctuating forces are required to comply
with the rules of quantum mechanics.

Detailed comparison of the properties of modes in these three
systems is provided in Fig.~\ref{fig5} by determining the maximal
values of the nonclassicality depths $ \tau $, $\tau_1 $, and $
\tau_2 $ and Fig.~\ref{fig6} by plotting the maximal values of the
negativity $ E_N $ and the steering parameters $ S_{1\rightarrow
2} $ and $ S_{2\rightarrow 1} $. 
We note that, in the passive PTSS, mode 2 is
asymptotically nonclassical and the corresponding parameters
including the nonclassicality depth $ \tau_2 $ are given in
Eqs.~(\ref{40}) and (\ref{41}).

\subsection{Comparison with passive $\mathcal{PT}$-symmetric system}

\begin{figure*}[t]  
 \centerline{ \includegraphics[width=0.98\hsize]{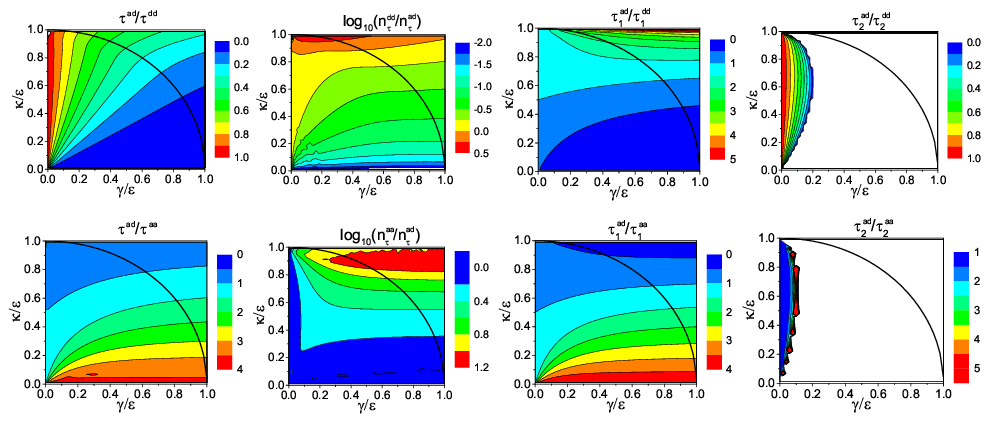} }
 \vspace{2mm}
 \centerline{ \small (a) \hspace{.22\hsize} (b) \hspace{0.22\hsize} (c) \hspace{0.22\hsize} (d)}
 \caption{ {(a) Nonclassicality depth $ \tau^{\rm ad} $ and (b) the corresponding mean photon
   number $ n_{\tau}^{\rm ad} $, (c) [(d)] local nonclassicality depth $ \tau_{1}^{\rm ad} $ [$ \tau_{2}^{\rm ad} $] of mode 1 [2]
   of standard PTSS relative to the values of passive PTSS
   with only doubled damping (superscript
   dd) and active PTSS with only doubled amplification (superscript aa)
   as they depend on model parameters
   $ \gamma/\epsilon $ and $ \kappa/\epsilon $.
   In white areas, $ \tau_{2}^{\rm ad} = 0 $.
   Solid black curves identify positions of EPs
   in PTSS as well as systems with only doubled damping and
   only doubled amplification.
   The superscript notation is explained in Fig.~\ref{fig1}.} }
 \label{fig5}
\end{figure*}

According to Figs.~\ref{fig5}(a-d, first row), the nonclassicality
depths $ \tau $ and $ \tau_2 $ are always smaller for the standard
PTSS compared to those of the passive one. This is true also for
the nonclassicality depth $ \tau_1 $ for $ \kappa/\epsilon $
smaller than approx. 0.6.
Greater values of $ \tau_1 $ for the standard PTSS than those for
the passive one are reached only for $ \kappa/\epsilon $ greater
than approx. 0.6. This is so, because doubled damping of mode 1
does not provide enough time for nonclassical-state generation for
greater nonlinear coupling constants $ \kappa/\epsilon $.
Moreover, in the standard PTSS, the amplified mode 2 results in
larger mode-2 amplitudes, which in turn enhance the effective
 {physical} nonlinearity. The standard PTSS also provides the
nonclassical states with greater overall intensities compared to
those of the passive PTSS, excluding the area of parameters with $
\kappa/\epsilon $ above 0.8 [see Fig.~\ref{fig5}(b, first row)].
The comparison of the behavior of quantum correlations described
by the negativity and steering parameters as presented in
Figs.~\ref{fig6}(a-d, first row) is even more straightforward. The
standard PTSS always gives smaller maximal values of the
negativity $ E_N $ and the steering parameters $ S_{1\rightarrow
2} $ and $ S_{2\rightarrow 1} $ and the states with smaller
overall intensities. In the passive PTSS, the states violating the
Bell inequalities are generated in the wide area of the system
parameters, similarly as in the damped part of the standard system
[compare Figs.~\ref{fig4}(d) and \ref{fig4}(a)]. These states also
attain greater overall intensities, as documented in
Fig.~\ref{fig4}(e). This contrasts with the behavior of the
standard PTSS that provides such states only for very small values
of the damping and amplification constants $ \gamma $. This
behavior qualitatively originates in the stronger detrimental
effects of the fluctuating forces in the amplified mode of the
standard PTSS compared to those of mode 2 of the passive system
(no amplification).

\subsection{Comparison with active $\mathcal{PT}$-symmetric
system}

Stronger fluctuating forces in the active PTSS imply in general
worse conditions for the nonclassical-state generation compared to
the standard PTSS. Indeed, the graphs in Figs.~\ref{fig5}(a-d,
second row) giving the maximal values of nonclassicality depths $
\tau $, $ \tau_1 $, and $ \tau_2 $ confirm this. Only when $
\kappa/\epsilon $ is greater than approx. 0.6 mode 1 in the active
system (no damping) attains greater nonclassicality depths $
\tau_1 $ compared to the damped mode 1 of the standard PTSS.
\begin{figure*}[t]  
 \centerline{\includegraphics[width=0.98\hsize]{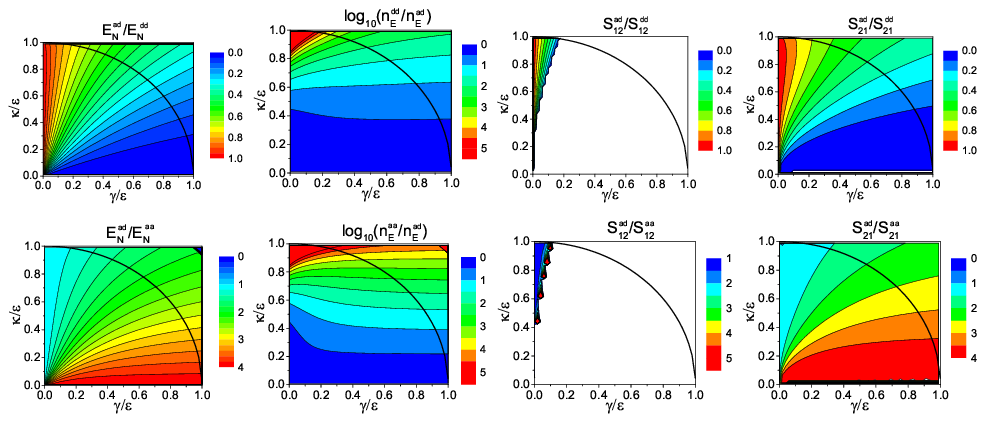}}
 \vspace{2mm}
 \centerline{ \small (a) \hspace{.22\hsize} (b) \hspace{0.22\hsize} (c) \hspace{.22\hsize} (d)}
 \caption{ {(a) Negativity $ E_N^{\rm ad} $ and (b) the corresponding mean photon
   number $ n_{E}^{\rm ad} $, (c) [(d)] steering parameter $ S_{1\rightarrow
   2}^{\rm ad} $ [$ S_{2\rightarrow 1}^{\rm ad} $]
   of standard PTSS relative to the values of passive PTSS (superscript
   dd), and active PTSS (superscript aa) as they depend on model parameters
   $ \gamma/\epsilon $ and $ \kappa/\epsilon $.
   In white areas, $ S_{1\rightarrow
   2}^{\rm ad} = 0 $. Solid black curves identify positions of EPs
   in PTSS as well as systems with doubled damping and
   amplification.
   The superscript notation is explained in Fig.~\ref{fig1}.} }
 \label{fig6}
\end{figure*}
This effect is due to the higher effective  {physical}
nonlinearity of the active PTSS, which arises from its larger mode
amplitudes. This behavior of mode 1 also results in greater
maximal values of the global nonclassicality depth $ \tau $ of the
active PTSS compared to those of the standard one in certain
subarea around $ \kappa/\epsilon =1 $. In general the active PTSS
gives greater overall intensities of the generated nonclassical
states. As for quantum correlations and opposed to what was
written when comparing with the passive PTSS, the standard PTSS
always gives greater maximal values of the negativity $ E_N $ as
well as the steering parameters $ S_{1\rightarrow 2} $ and $
S_{2\rightarrow 1} $. However the nonclassical states attain
smaller overall intensities than those reached in the active PTSS
[see Fig.~\ref{fig6}(b, second row)]. The ability of both systems
to generate the states with the Bell nonlocality is very weak and
it is restricted to very low values of the damping and
amplification constants $ \gamma/\epsilon $.

We note that, in the above calculations, we assumed the reservoir
two-level atoms in the ground states to consistently describe
damping and in the excited states to consistently describe
amplification. Nevertheless, the two-level reservoir atoms for
damping can partly be in their excited states similarly as the
reservoir two-level atoms for amplification can partly be in their
ground states. These modifications of the reservoir properties
make closer the behavior of the above considered standard,
passive, and active PTSSs. In the asymptotic limit of equally
populated ground and excited levels of the atoms in both
reservoirs, the behavior of all PTSSs is identical.

\section{Quantumness and hierarchy of quantum correlations}

The results presented above about the system's nonclassicality and
its quantum correlations, made systematically across the entire
hierarchy of quantum correlations (entanglement, steering, and
Bell nonlocality), show that the balance between damping and
amplification in the standard PTSS, and the ensuing specific
system dynamics, do not improve, except for minor cases, the
system's ability to generate nonclassical states exhibiting
different kinds of quantumness. We note that similar hierarchies
of quantumness potentials were studied in
Refs.~\cite{Kadlec2024,Kadlec2024a}, in the context of
single-qubit states.

In general, the set of states exhibiting the Bell nonlocality
forms a subset of the set of steerable states. Similarly,
steerable states constitute a subset of entangled states. Finally,
all entangled states belong to the class of nonclassical states,
together with those exhibiting only local nonclassicality. This
hierarchy of quantum states, classified according to different
forms of quantumness, is also preserved for the quantities
discussed above, which are defined as the maxima taken over the
dimensionless time $\epsilon t$.

Moreover, we find that states exhibiting the Bell nonlocality---as
the states with the strongest nonclassicality---are effectively
generated only in systems without amplification, across broad
ranges of the system parameters [see Fig.~\ref{fig4}].
\begin{figure*}[t]  
 \centerline{ \includegraphics[width=0.8\hsize]{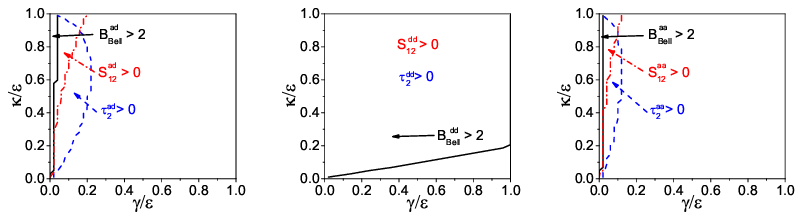}}
 \vspace{2mm}
 \centerline{ \small (a) \hspace{.22\hsize} (b) \hspace{0.22\hsize} (c)}
 \vspace{2mm}
 \centerline{ \includegraphics[width=0.6\hsize]{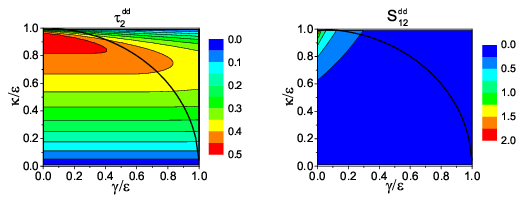} }
 \vspace{2mm}
 \centerline{ \small (d) \hspace{.22\hsize} (e)}
 \caption{Nonclassical boundaries for Bell parameter $ B_{\rm Bell} > 2 $ (black solid
  curve), local nonclassicality depth $ \tau_2 > 0 $ of mode 2 (blue dashed curves), and
  steering parameter $ S_{1\rightarrow 2} > 0 $ (red dashed-dotted curves) in the
  plane of model parameters $ \gamma/\epsilon $ and $ \kappa/\epsilon
  $ drawn for (a) standard, (b) passive, and (c) active PTSSs. In
  (b), $ \tau_2 > 0 $ and $ S_{1\rightarrow 2} > 0 $ hold for all model
  parameters. (d) Local nonclassicality depth $ \tau_2^{\rm dd} $
  of mode 2 and (e) steering parameter $ S_{1\rightarrow 2}^{\rm dd} $ for passive PTSS
  are shown as they depend on model parameters $ \gamma/\epsilon $ and $ \kappa/\epsilon
  $; solid black curves identify positions of EPs in passive PTSS.
  The superscript notation is explained in Fig.~\ref{fig1}.}
 \label{fig7}
\end{figure*}
For comparison, these ranges are plotted in Figs.~\ref{fig7}(a),
\ref{fig7}(b), and \ref{fig7}(c) for the standard, passive, and
active PTSSs, respectively. Even weak amplification prevents the
system from generating the Bell-nonlocal states. In contrast,
steerable and entangled states are observed in standard, passive,
and active PTSSs for any values of the system parameters. Steering
is, however, strongly asymmetric. In the system with
amplification, the non-amplified mode is steerable for any value
of the system parameters, but the amplified mode is steerable only
for very weak values of the considered damping or amplification
[see Figs.~\ref{fig3}(c) and \ref{fig6}(c); compare the boundaries
in Figs.~\ref{fig7}(a)--\ref{fig7}(c); see also
Fig.~\ref{fig7}(e)].

Similarly, the non-amplified mode exhibits local nonclassicality
for any values of the system parameters. This is not the case for
observing local nonclassicality in the amplified mode, which
requires smaller values of the considered damping or amplification
constants [see Figs.~\ref{fig2}(d) and \ref{fig5}(d); compare the
boundaries in Figs.~\ref{fig7}(a)--\ref{fig7}(c); see also
Fig.~\ref{fig7}(d)].

In the case of the standard PTSS, the states violating the Bell
inequalities, the states in which the amplified mode exhibits
local nonclassicality, and the states allowing steering of the
amplified mode cannot be generated across the entire range of
system parameters. These states are reached only for small values
of the damping and amplification constants. In contrast, entangled
and steerable states are easily obtained throughout the full
parameter space.


\section{Evolution of the negativity, nonclassicality depth and their generation speed}

\begin{figure*}[t]  
 \centerline{ \includegraphics[width=0.98\hsize]{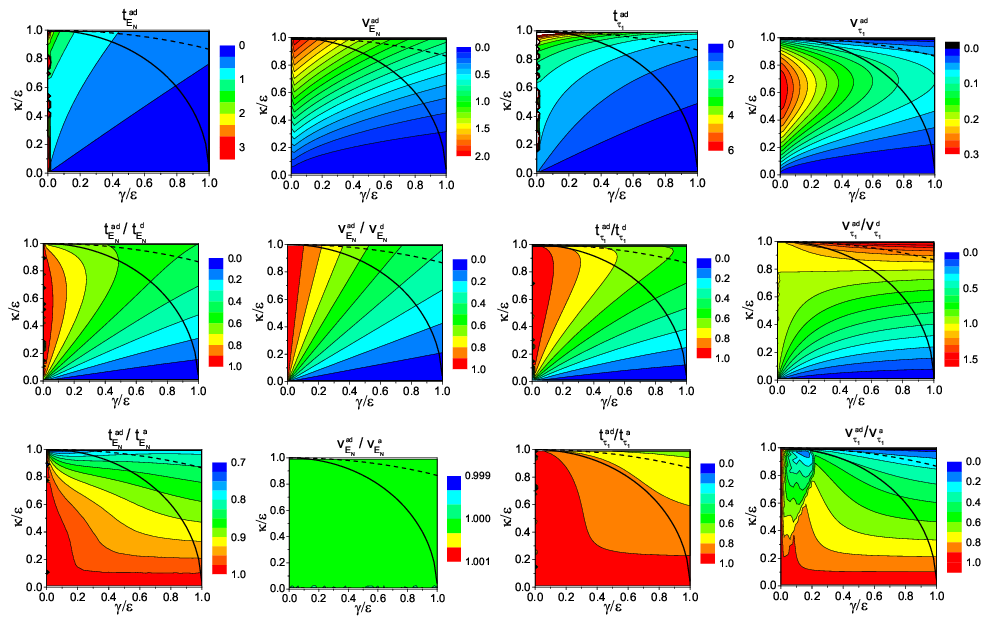} }
 \vspace{2mm}
 \centerline{ \small (a) \hspace{.225\hsize} (b) \hspace{.225\hsize} (c) \hspace{.225\hsize} (d) }
 \caption{(a) Time instant $ t^{\rm ad}_{E_N} $ corresponding to the maximal negativity $ E_N $, (b) maximal
  speed $ v^{\rm ad}_{E_N} $ of the negativity generation, (c) time instant $ t^{\rm ad}_{\tau_1} $ yielding the maximal nonclassicality depth
   $ \tau_1 $ of mode 1, and (d) maximal speed $ v^{\rm ad}_{\tau_1} $ of nonclassicality-depth $ \tau_1 $ generation
   in standard PTSS as they depend on model parameters
   $ \gamma/\epsilon $ and $ \kappa/\epsilon $. The values of the
   drawn parameters are compared with those originating in the
   model with considered only damping (superscript d) and only
   amplification (superscript a). Solid [dashed] black curves identify positions of EPs
   in PTSS [systems with only damping and only amplification].
   The superscript notation is explained in Fig.~\ref{fig1}.}
 \label{fig8}
\end{figure*}

Above we have compared the maximal values of nonclassicality
depths and several quantifiers of quantum correlations attained
during the system evolution. Here, we address the process of
nonclassical-state generation in a more detailed way by analyzing
the times needed to arrive at these maximal values and speeds of
their generation in the standard PTSS. As typical examples, we
investigate in Figs.~\ref{fig8}(a-d) the times $ t $ and speeds $
v $ belonging to the negativity $ E_N $ and local nonclassicality
depth $ \tau_1 $ of mode 1. We can see in Figs.~\ref{fig8}(a-d,
first row), that greater values of the damping and amplification
constants $ \gamma $ shorten these times $ t^{\rm ad}_{E_N} $ and
$ t^{\rm ad}_{\tau_1} $ and also slow down the negativity (speed $
v^{\rm ad}_{E_N} $) and nonclassicality (speed $ v^{\rm
ad}_{\tau_1} $) generation. Contrary to this, greater values of
the nonlinear coupling constant $ \kappa/\epsilon $ make the times
$ t^{\rm ad}_{E_N} $ and $ t^{\rm ad}_{\tau_1} $ longer and the
speed $ v^{\rm ad}_{E_N} $ of negativity generation faster. On the
other hand, the maximal speeds $ v^{\rm ad}_{\tau_1} $ of mode-1
nonclassicality generation are reached for $ \kappa/\epsilon
\approx 0.6 $. This is caused by the interplay of the increasing
ability to generate nonclassical states and slowing down the
system evolution when moving towards an EP with the increasing
nonlinear constant $ \kappa/\epsilon $ \cite{PerinaJr2023}.

Comparing the behavior of the standard PTSS with the systems with
only damping [see Figs.~\ref{fig8}(a-d, second row)] and only
amplification [see Figs.~\ref{fig8}(a-d, third row)], the maximal
values of negativity $ E_N $ and local nonclassicality $ \tau_1 $
are always reached faster in the standard PTSS. The maximal speeds
$ v^{\rm ad}_{E_N} $ of negativity generation are always faster
for the system with only damping which accords with greater
maximal values of the negativity $ E_N $ reached in this system.
Interestingly, the maximal speeds $ v^{\rm ad}_{E_N} $ are very
close in the standard PTSS and the system with only amplification,
as documented in Fig.~\ref{fig8}(b, third row). The maximal speed
$ v^{\rm ad}_{\tau_1} $ of mode-1 nonclassicality generation in
the standard PTSS is usually smaller than those of the systems
with only damping and only amplification. Only when $
\kappa/\epsilon \ge 0.8 $, amplification in mode 2 of the standard
system and the resulting larger amplitudes of the mode) allows for
faster mode-1 nonclassicality generation compared to the system
with only damping Fig.~\ref{fig8}(d, second row).

The generally lower maximum values of nonclassicality depths and
the analyzed quantum-correlation quantifiers observed in the
standard PTSS---compared to systems with only damping or only
amplification---can be attributed to two key factors: a shorter
duration of nonclassical-state generation and a slower buildup of
nonclassical properties.
This qualitatively resembles the behavior of fields in nonlinear
three-mode parametric processes with phase mismatch
\cite{Boyd2003}.

The above findings, which identify the passive PTSS as the most
efficient source of nonclassical states, support its practical
implementation---especially since achieving damping is generally
simpler than achieving amplification. The realization requires a
nonlinear medium with a strong $ \chi^{(2)} $ nonlinearity
\cite{Boyd2003}. When such a medium is placed inside a resonator,
the resonator enhances the effective nonlinearity and
simultaneously introduces damping to both frequency down-converted
modes through leakage via the resonator mirrors. Adding a
birefringent material to the resonator, which linearly couples the
modes, then completes the formation of a passive PTSS. It is worth
noting that nonlinear $ \chi^{(2)} $ crystals inside resonators
have been reliable sources of squeezed light for decades
\cite{Mehmet2011,Lvovsky2009}.

We note that realizing an active PTSS is also feasible, for
example, using spin-polarized laser technology. In these lasers,
two optical modes coexist, each pumped by electrons polarized in
orthogonal spin directions (left-to-right and right-to-left). The
natural coupling between these modes in various photonic
heterostructures enables the formation of an active PTSS
\cite{Drong2020}. In Refs.~\cite{Drong2021,Drong2022}, such
spin-polarized lasers featuring exceptional points were analyzed,
with field saturation effectively acting as a Kerr-type
nonlinearity.

\section{Conclusions}

Numerical analysis of  {two bosonic modes coupled linearly and by
parametric down-conversion performed} across the full range of
system parameters reveals that the standard
$\mathcal{PT}$-symmetric system (PTSS)---characterized by balanced
damping and amplification---does not, in general, enhance the
system's ability to generate nonclassical states of light. This
conclusion is supported across the entire hierarchy of
nonclassicality and quantum correlations by comparing in turn the
nonclassicality depths, negativity, steering parameters, and the
Bell parameter for the standard PTSSs and related systems affected
solely by either damping or amplification.

While the standard PTSS outperforms its active counterpart---where
one mode is undamped and the other is doubly amplified---it is
clearly outperformed by the passive variant, in which one mode is
doubly damped and the other is unaffected by amplification. This
is particularly notable because all three systems possess
identical eigenvectors and real parts of their eigenfrequencies.
Their differing behavior arises from distinct characteristics of
the fluctuating forces accompanying damping and amplification
(involving spontaneous emission).

The reduced ability of the standard PTSS to generate nonclassical
states can be attributed to two factors: a shorter time window
over which nonclassical properties arise, and a slower rate at
which these properties develop.

In conclusion, the most significant benefits of the PTSS dynamics
in  {physically} nonlinear systems are realized in the passive
PTSSs configuration. Such systems can generate highly nonclassical
states---featuring entanglement, quantum steering, and Bell
nonlocality---across wide parameter ranges. They also allow for
practical experimental realizations based on parametric
down-conversion in crystals embedded into optical resonators.

\acknowledgments J.P. acknowledges support by the project No.
25-15775S of the Czech Science Foundation. K.B., G.C., A.K.-K.,
and A.M. were supported by the Polish National Science Centre
(NCN) under the Maestro Grant No. DEC-2019/34/A/ST2/00081. J.K.K.
and W.L. acknowledge the support provided by the program of the
Polish Ministry of Science entitled 'Regional Excellence
Initiative', project no. RID/SP/0050/2024/1. J.P. also
acknowledges support by the project ITI
CZ.02.01.01/00/23\_021/0008790 of the Ministry of Education,
Youth, and Sports of the Czech Republic and EU.


%

\end{document}